\documentclass[
twocolumn,
amsmath,amssymb,
aps,
pre,
]{revtex4-2}

\usepackage{graphicx}
\usepackage{dcolumn}
\usepackage{bm}
\usepackage{placeins}
\usepackage[dvipsnames]{xcolor}

\usepackage[normalem]{ulem}
 
\begin{document}

\title{Effects of large-scale advection and small-scale turbulent diffusion\\ on vertical phytoplankton dynamics }

\author{Vinicius Beltram Tergolina}
 \thanks{Corresponding author}\email{vinicius.beltram-tergolina@univ-lille.fr}

  \affiliation{Univ. Lille, ULR 7512 - Unit\'e de M\'ecanique de Lille Joseph Boussinesq (UML), F-59000 Lille, France}
\author{Enrico Calzavarini}%

  \affiliation{Univ. Lille, ULR 7512 - Unit\'e de M\'ecanique de Lille Joseph Boussinesq (UML), F-59000 Lille, France}
\author{Gilmar Mompean}

  \affiliation{Univ. Lille, ULR 7512 - Unit\'e de M\'ecanique de Lille Joseph Boussinesq (UML), F-59000 Lille, France}
 \author{Stefano Berti}%

  \affiliation{Univ. Lille, ULR 7512 - Unit\'e de M\'ecanique de Lille Joseph Boussinesq (UML), F-59000 Lille, France}

\date{\today}

\begin{abstract}

 Turbulence has been recognized as a factor of paramount importance for the survival or extinction 
 of sinking phytoplankton species. However, dealing with its multiscale nature in models of coupled fluid and biological dynamics is a formidable challenge. Advection by coherent structures, as those related to winter convection and Langmuir circulation, is also recognized to play a role in the survival and localization of phytoplankton. In this work we revisit a theoretically appealing model for phytoplankton vertical dynamics, and numerically investigate how large-scale fluid motions affect the survival conditions and the spatial distribution of the biological population. For this purpose, and to work with realistic parameter values, we adopt a kinematic flow field to account for the different spatial and temporal scales of turbulent motions. The dynamics of the population density are described by an advection-reaction-diffusion model with a spatially heterogeneous growth term proportional to sunlight availability. We explore the role of 
 fluid transport by progressively increasing the complexity of the flow in terms of spatial and temporal scales. We find that, due to the large-scale circulation, phytoplankton accumulates in downwelling regions and its growth is reduced, confirming previous indications in slightly different conditions. We then explain the observed phenomenology in terms of a plankton filament model. 
 Moreover, by contrasting the results in our different flow cases, we show that the large-scale 
 coherent structures have an overwhelming importance. Indeed, we find  that smaller-scale motions only quite weakly affect the  dynamics, without altering the general mechanism identified. 
 Such results are relevant for parameterizations in numerical models of phytoplankton life cycles in realistic oceanic flow  conditions.

\end{abstract}

\maketitle

\section{\label{sec:intro} Introduction}
The occurrence of phytoplankton blooms is a topic of considerable interest to oceanography, given its relation to primary production and carbon export~\cite{MLbook,WFbook}. The understanding of the biological and physical conditions leading to blooms is, however, still incomplete. This is due to the variety of intervening processes, as well as to the lack of detailed information about the vertical structure of the phytoplankton biomass distribution, and of the fluid flows that shape it.

Modeling studies in the field have been useful to rationalize the evidences from experimental observations into theoretical, predictive, frameworks~\cite{huisman1999critical,okubo2001diffusion,ryabov2008population,ryabov2010vertical}. Among these theories, those addressing vertical dynamics in light-limited environments have a rich history, starting from the introduction of the concept of critical depth by Gran and Braarud~\cite{gran1935quantitative}, which lead to  Sverdrup's celebrated critical depth hypothesis~\cite{sverdrup1953conditions}. 
The idea is that phytoplankton blooms would only be possible when the mixed layer, the weakly stratified upper part of the water column, is shallower than a certain critical depth, defined as the point where the population depth-integrated gains (due to photosynthesis) surpass the depth-integrated losses (due to grazing and respiration). Sverdrup's reasoning relies on several assumptions: a well-mixed fluid layer; negligible nutrient limitations; direct proportionality between the photosynthetic biomass production and the available energy from the incoming radiation; a constant light attenuation coefficient throughout the water column.
Building on such ideas, and aiming to improve realism, subsequent studies started to address also the role of algal self-shading~\cite{shigesada1981analysis} and of turbulence~\cite{okubo2001diffusion} on phytoplankton life cycles. A unifying framework for different previous theories invoking the importance of the water-column depth and of turbulence intensity emerged from the influential work of Huisman and collaborators on sinking phytoplankton dynamics~\cite{huisman2002sinking,huisman2002population}. 
Such previous studies, however, focused on the one-dimensional (1D) dynamics along the vertical, assuming that turbulence can be approximated by a diffusive process. Therefore, they could not account for the effects due to its complex, multiscale, character. In addition, persistent and well organized two-dimensional (2D) fluid motions, as those characterizing winter convection, were also suggested to play an important role on phytoplankton survival~\cite{backhaus1999phyto,backhaus2003convection}.

In this work we develop a 2D model that allows us to include the effects of both large-scale fluid motions and smaller scale turbulent ones on the survival dynamics of sinking phytoplankton in light-limited environments, with the aim of extending the picture drawn from simpler 1D models~\cite{huisman2002sinking,huisman2002population}. 
In fact, studies discussing the influence of turbulence and horizontal advection over plankton cycles, patchiness and survival have already shown that fluid transport and mixing may considerably impact the evolution of the population distribution (see, e.g.,~\cite{koszalka2007plankton,bracco2009horizontal}). Our approach shares some similarity with the numerical investigations reported in Ref.~\cite{taylor2011shutdown}, based on large-eddy simulations (LES), and in Ref.~\cite{lindemann2017dynamics}, employing a kinematic model of a stationary flow. 
In the latter works, however, it is less evident how to disentangle the contributions from large and small flow scales than in ours, where we proceed incrementally, 
adding time dependency and smaller scales to the flow separately. 

More specifically, we carry out numerical simulations of an advection-reaction-diffusion model for the phytoplankton population density, in which the advecting velocity field is specified by a kinematic cellular flow. The flow will consist of a single (steady or unsteady) large-scale mode spanning the whole depth of the system, and a series of temporally varying modes with smaller and smaller length scales. 
Such a choice allows for a simplified description of the buoyancy and wind-driven flows~\cite{thorpe2005turbulent}, such as convective currents and Langmuir circulations, often encountered in the upper layers of oceans and lakes. A distinguished feature of these flows is, indeed, the simultaneous presence of (small-scale) turbulence and (large-scale) coherent structures. 

We investigate the model dynamics, as a function of the flow intensity and spatiotemporal structure, in a vertical fluid layer of fixed depth. Our system can then be thought as representative of a coastal area or a lake, where the mixed-layer depth undergoes smaller variations than in the open ocean. Interestingly, some studies motivated by either environmental monitoring~\cite{moreno2009influence} or the aim of testing different theoretical hypotheses~\cite{behrenfeld2010abandoning,mignot2016spring,mignot2018floats} point to the relevance of such fluid environments. Note, also, that, as our main goal is to focus on the interplay between fluid transport and biological growth, for the sake of simplicity, we neglect the dynamics of nutrients and we do not explicitly represent those of predators. In our setup, light availability is thus the only limiting factor for phytoplankton growth.

This article is organized as follows. We introduce the model dynamics for the pytoplankton density, and the kinematic flow field, in Sec.~\ref{sec:model}. 
The numerical results are reported in Sec.~\ref{sec:results}, where we separately discuss the different flow cases considered. Finally, discussions and conclusions are presented in Sec.~\ref{sec:conclus}.

%%%%%%%%%%%%%%%%%%%%%%%%%%%%%%%%%%%%%%%%%%%%%%%%%%
\section{\label{sec:model} Model}

We adopt a 2D advection-reaction-diffusion model 
for the dynamics of the population density field $\theta(x,z,t)$ (number of individuals per unit volume), whose evolution equation reads:
\begin{equation}
    \frac{\partial \theta}{\partial t} = \left[ p(I) - l \right]\theta - \bm{v} \cdot \bm{\nabla} \theta + D \bm{\nabla}^{2}\theta .
    \label{full}
\end{equation}
We consider such dynamics in a  vertical fluid layer, intended to represent the mixed layer, of horizontal and vertical sizes $L_x$ and $L_z$, respectively, with rigid walls at the top and bottom boundaries.

Biological growth is controlled by a production rate, $p$, and a loss rate, $l$. Advection is realized by a 2D incompressible flow $\bm{u}=(u_x,u_z)$ and phytoplankton is assumed to sink with a speed $v_{sink}\hat{\bm{z}}$, where $\hat{\bm{z}}$ is the unitary vector pointing downward in the vertical direction; the total velocity appearing in Eq.~(\ref{full}) is thus $\bm{v}=\bm{u}+v_{sink}\hat{\bm{z}}$. The coefficient $D$ represents an effective diffusivity, due to both small-scale unresolved turbulent motions and possible swimming behavior. The production term accounts for both water background turbidity, with coefficient $\kappa_{bg}$, and population self-shading, with an attenuation factor $\kappa$. Its functional form is:
\begin{equation}
    \label{eq_prod}
    p(I) = \frac{p_{max}I}{H+I},
\end{equation}
where $p_{max}$ is the maximum specific production rate, $H$ is a half-saturation constant and the time- and depth-dependent light-intensity is expressed as follows, according to Lambert-Beer's law:
\begin{equation}
    I(z,t) = I_{in}e^{-\int_{0}^{z}\kappa \theta(s , t)ds - \kappa_{bg}z},
\end{equation}
with $I_{in}$ the incident light (at the surface, where $z=0$). 
The biological parameter values adopted in our study, representative of realistic situations, are reported in Table \ref{tab:table1}. They are extracted from~\cite{huisman2002sinking}, with growth parameters measured for freshwater phytoplankton species and $\kappa_{bg}$ for clear lakes and coastal areas~\cite{huisman2002population}. 

\begin{table*}
\caption{\label{tab:table1} 
Parameters of the biological dynamics.}
\begin{ruledtabular}
\begin{tabular}{ccc}
 \textbf{Parameter}& \textbf{Value}& \textbf{Meaning}\\ \hline
 $\kappa_{bg}$   & $0.2$~m$^{-1}$ & Background turbidity\\
 $\kappa$ &   $1.5 \cdot 10^{-11}$~m$^2$~cell$^{-1}$  & Specific light attenuation of phytoplankton\\
 $p_{max}$ & $0.04$~h$^{-1}$ & Maximal specific production rate\\
 $l$ & $0.01$~h$^{-1}$ & Specific loss rate\\
 $H$    & $30\,\mu$mol~photons~m$^{-2}$~s$^{-1}$ &  Half-saturation constant of light-limited growth\\
 $I_{in}$&   $350\,\mu$mol~photons~m$^{-2}$~s$^{-1}$  &Incident light intensity\\
 $v_{sink}$& $0.04$~m~h$^{-1}$  &Phytoplankton sinking velocity\\
\end{tabular}
\end{ruledtabular}
\end{table*}

We consider advection by a prescribed cellular flow, which is intended to model the presence of eddying fluid motion on different scales. The velocity field is then obtained as $\bm{u}=\left(-\partial_z \Psi, \partial_x \Psi\right)$ from a streamfunction that, in the general form (see also~\cite{solomon1988chaotic,boffetta2000nonasymptotic,lacorata2017chaotic,lacorata2019fsle}), can be written as:
\begin{equation}
    \Psi(x,z,t) = \Psi_{L}(x,z,t) + \Psi_{s}(x,z,t),
    \label{flowcomplt}
\end{equation}
where
\begin{equation}
    \Psi_{L}(x,z,t) = -\frac{U_{1}}{k_{1}}\sin\left \{k_{1}[x-s_{1}\sin(\omega_{1}t)]\right\} \sin(k_{1}z),
    \label{flowlarge}
\end{equation}
and
\begin{equation}
    \begin{split}
        \Psi_{s}(x,z,t) & = -\sum_{i = 2}^{n_{k}} \frac{U_{i}}{k_{i}}\sin\left \{k_{i}[x-s_{i}\sin(\omega_{i}t)]\right\} \\
  &  \times \sin\{k_{i}[z-\beta(z)s_{i}\sin(\omega_{i}t)]\}.
      \label{flowsmall}
    \end{split}
\end{equation}
Here $\Psi_{L}$ represents a large scale persistent structure that is allowed to oscillate in the horizontal and $\Psi_{s}$ represents smaller-scale vortices that oscillate both in the horizontal and in the vertical. In Eq.~(\ref{flowsmall}), $n_k$ is the number of modes selected, $L_{i} = L_{1}\eta^{1-i}$ is the typical length scale of mode $i$, with $\eta>1$ a scale separation factor and $L_1=L_x$ the largest flow scale,  
$k_{i} = 2\pi/L_{i}$ the corresponding wavenumber, and $U_i$ the typical flow intensity at scale $L_i$ [the same notation is used in Eq.~(\ref{flowlarge}), where $i=1$]. We choose to account for a possible explicit time dependency of the flow field in the form of oscillations with amplitudes $s_i$ and pulsations $\omega_i$. To respect no-flux boundary conditions for Eq.~(\ref{full}) in $z=0$ and $z=L_z$, 
$\left[ v_{sink} \theta-D \partial_z \theta \right]_{z=0,L_z}= 0 $, we use the function:
\begin{equation}
    \label{eq:dump}
    \beta(z) = \frac{1}{2} \left[ 
    \tanh\left(\frac{z - z_1}{\xi}\right) - 
    \tanh\left(\frac{z - z_2}{\xi}\right) \right],
\end{equation}
to damp vertical oscillations near the vertical boundaries and therefore guarantee that $u_{z}$ 
is negligibly small there. A similar choice of a damping function was adopted to study chemical reactions in closed vessels~\cite{lopez2002efficiency}. In our case, the parameter values $z_{1} = 2$~m, $z_{2} = L_{z} - z_{1}$ and $\xi=1$~m turned out to be adequate for this purpose. In the following we will consider velocity fields with increasing degree of complexity, namely a steady  one-mode flow (\ref{sec:results steady}), an unsteady one-mode flow (\ref{sec:results cell}), and a multiscale time-dependent flow (\ref{sec:results multiscale}) . While our main focus will be on the interplay between large-scale advection and small-scale turbulent diffusion, with this choice we aim at exploring the effect of smaller temporal and spatial scales on the biological dynamics. 

We numerically integrate Eq.~(\ref{full}) by means of a pseudo-Lagrangian algorithm~\cite{abel2001front,berti2005mixing,sandulescu2007plankton,GF2020} (see the Appendix~\ref{sec:appendixa} for more details) in our rectangular domain with $L_x=2L_z$, using periodic and no-flux (as in~\cite{huisman2002sinking}) boundary conditions along the horizontal ($x$) and the vertical ($z$), respectively. The initial condition is a low uniform population density  [$\theta(t=0) = 5.5 \cdot 10^{6}$~cells~m$^{-3}$], but we checked in some selected cases that the results do not appreciably change if the population is initially present only in a small localized patch. To analyze the blooming conditions we mainly rely on the temporal behavior of the average biomass density, 
\begin{equation}
    \langle \theta \rangle(t) = \frac{1}{L_{x}L_{z}} \int_{0}^{L_{x}}\int_{0}^{L_{z}}\theta(x,z,t) \, dx dz,
    \label{eq:biomass}
\end{equation}
and the per-capita growth rate~(see, e.g., \cite{murray2007mathematical}),
\begin{equation}
    r_p(t)= \frac{1}{\langle \theta \rangle}\frac{\partial \langle \theta \rangle}{\partial t}.
\label{eq:pc_gr}
\end{equation}
In particular, after an initial transient, the latter quantity is expected to attain a statistically constant value $r_p$, corresponding to exponential growth ($r_p>0$) or decay ($r_p<0$) in the early regime before the onset of nonlinear dynamical effects (due to self-shading).  We also use $\overline{r}_{p}$ to indicate the time average of $r_{p}(t)$ over the entire simulation. Note that in a simulation of duration $T$ such time average can be expressed as $\overline{r}_{p} = (1/T) \ln{\left[\langle \theta \rangle(T)/\langle \theta \rangle(0)\right]}$.

%%%%%%%%%%%%%%%%%%%%%%%%%%%%%%%%%%%%%%%%%%%%%%%%%%
\section{\label{sec:results} Results}

%%%%%%%%%%%%%%%%%%%%%%%%%%%%%%%%%%%%%%%%%%%%%%%%%%
\subsection{Steady Flow \label{sec:results steady} }

\begin{figure}[h]
\includegraphics[scale = 0.8]{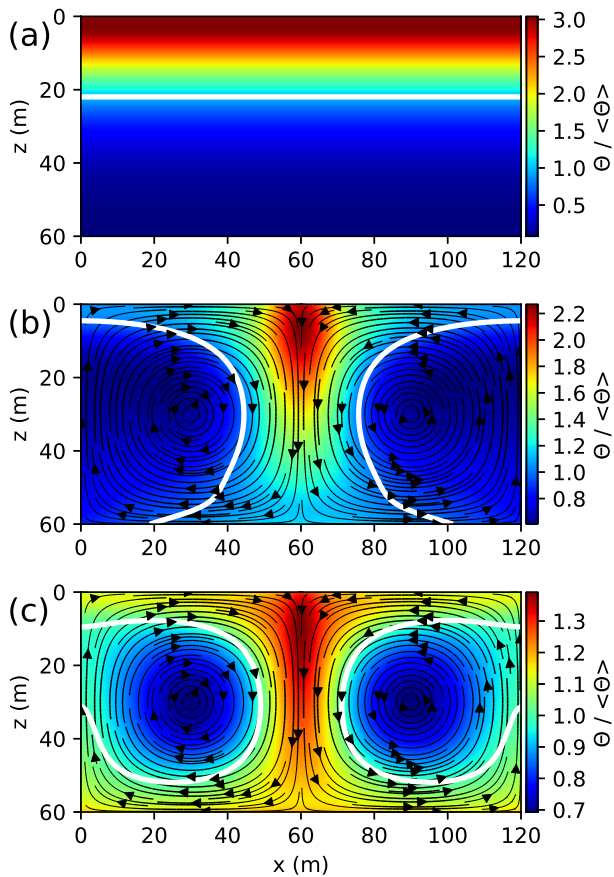}
\caption{\label{snapfieldsteady} 
Instantaneous normalized population density field $\theta(x,z,t^{*})/\langle \theta \rangle$ at a fixed instant of time $t^{*}=640$~h for $D=5$~cm$^2$~s$^{-1}$ and $U=(0,1.24,4.93)$~m~h$^{-1}$ 
[panels (a), (b) and (c), respectively], where $\langle \theta \rangle$ stands for the spatial average. The white line is the isoline $\theta/\langle \theta \rangle = 1$,
and $t^{*}$ is in the regime of stationary per-capita growth rate ($r_p(t)=r_p=\mathrm{const}$). The solid black lines in (b) and (c) represent flow streamlines, with arrows indicating the circulation direction.
}
\end{figure}

In the absence of a flow field ($\Psi=0$), our 2D model is equivalent to the original 1D one~\cite{huisman2002sinking} and numerical simulations reproduce the results of the latter, as verified by computing vertical population profiles, as well as the phase diagram summarizing the survival (or extinction) conditions versus the diffusivity $D$ and water-column depth (results not shown). A typical snapshot of the population density field is shown in Fig.~\ref{snapfieldsteady}(a), which clearly shows the independence of the $\theta$ field on the lateral direction $x$. 

A relevant feature of the original model is the existence of a turbulence window allowing for phytoplankton bloom, for large enough system depths. Determining analytical expressions for the critical conditions for population survival (i.e. blooming) or extinction is not an easy task, even in such a simple model~\cite{huisman2002sinking}. This difficulty is due to the heterogeneity of the environment and is common to different population dynamics' models (see, e.g.,~\cite{SG2001,ryabov2008population,vergni2012invasions} for other 1D systems). Adopting some simplifying assumptions, it is possible to obtain an approximate estimate of the minimum turbulent diffusivity (the lower bound of the turbulence window) required to compensate the sinking of phytoplankton, and hence to let the population survive~\cite{Riley_etal_1949,shigesada1981analysis,SG2001,huisman2002sinking}. Nevertheless, for the maximum turbulent diffusivity (the upper bound of the window), beyond which the population cannot outgrow the turbulent mixing rate to sustain the bloom in the upper part of the water column, no simple analytical expression is known~\cite{ebert2001critical,huisman2002sinking}.  

Here, we numerically investigate the effect of a large-scale steady cellular flow on the dynamics of the phytoplankton population and its  
survival/extinction transitions. 
The streamfunction corresponding to such a velocity field is
\begin{equation}
    \Psi_{L}^{st}(x,z) = -\frac{U}{k}\sin(kx)\sin(kz)
    \label{convect}
\end{equation}
i.e.~Eq.~(\ref{flowlarge}) where no explicit time dependency is included (with $k=k_1=\pi/L_z$ and $U = U_1$).  We consider a depth for which the turbulence window exists for the no-flow system ($L_z \gtrsim 60$~m), as documented in~\cite{huisman2002sinking}, and we fix the turbulent diffusivity to a value that is intermediate between the minimum ($D \approx 0.1$~cm$^2$~s$^{-1}$) and maximum $D \approx 100$~cm$^2$~s$^{-1}$) critical ones for blooming. Due to the increased computational times of simulations in larger spatial domains, we choose a depth value close to the minimum possible one, namely $L_z=60$~m. Streamlines corresponding to the flow from Eq.~(\ref{convect}) can be seen in Figs.~\ref{snapfieldsteady}(b,c). From these figures it is also evident that the flow impacts the spatial distribution of the population, which is no longer laterally homogeneous. We will discuss in more detail this point later in this section. 
%%%%%%%%%%%%%%%%%%
% FIG. 2
%%%%%%%%%%%%%%%%%%

\begin{figure}[h!]
\includegraphics[scale = 0.41]{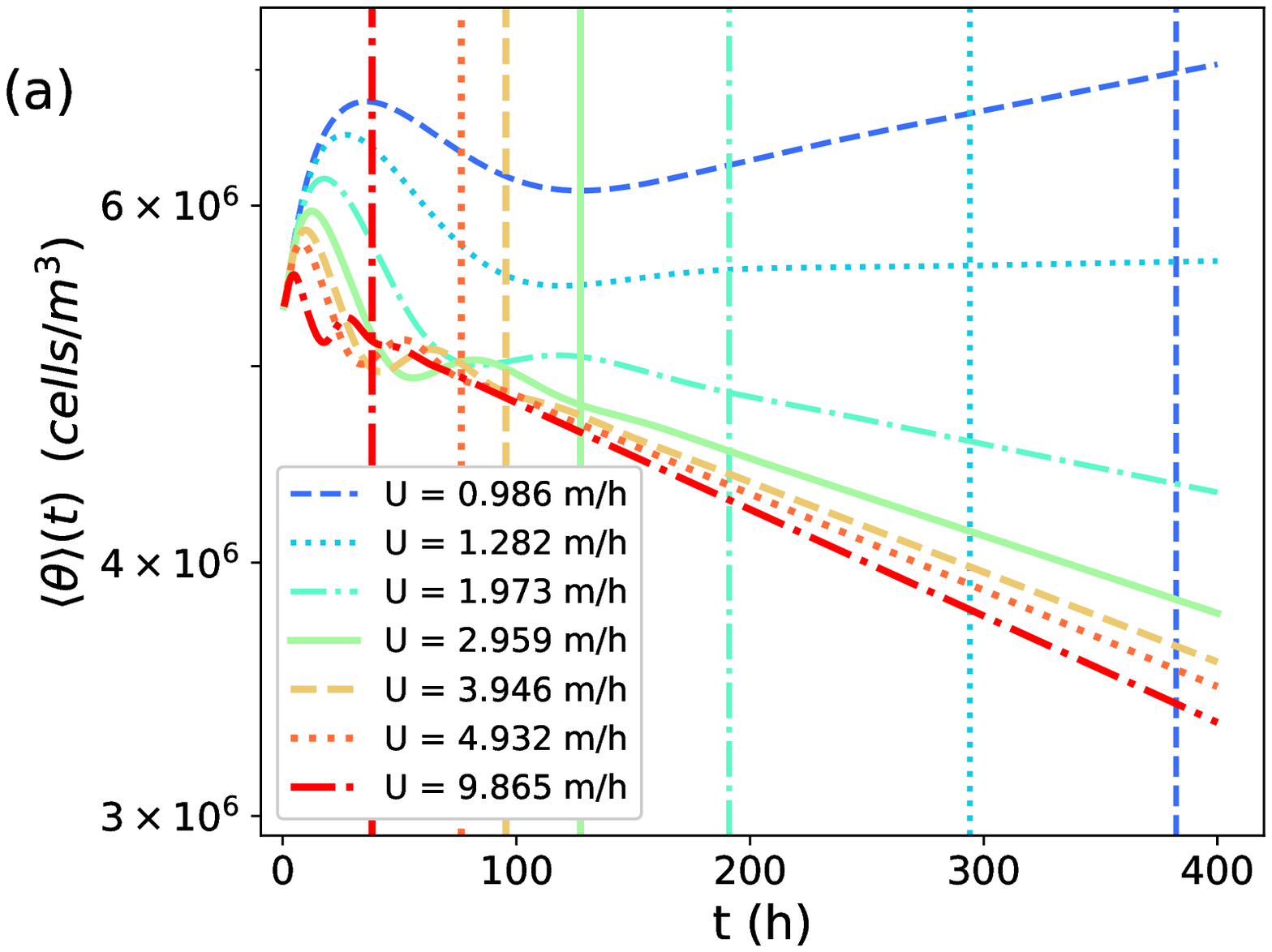}
\includegraphics[scale = 0.34]{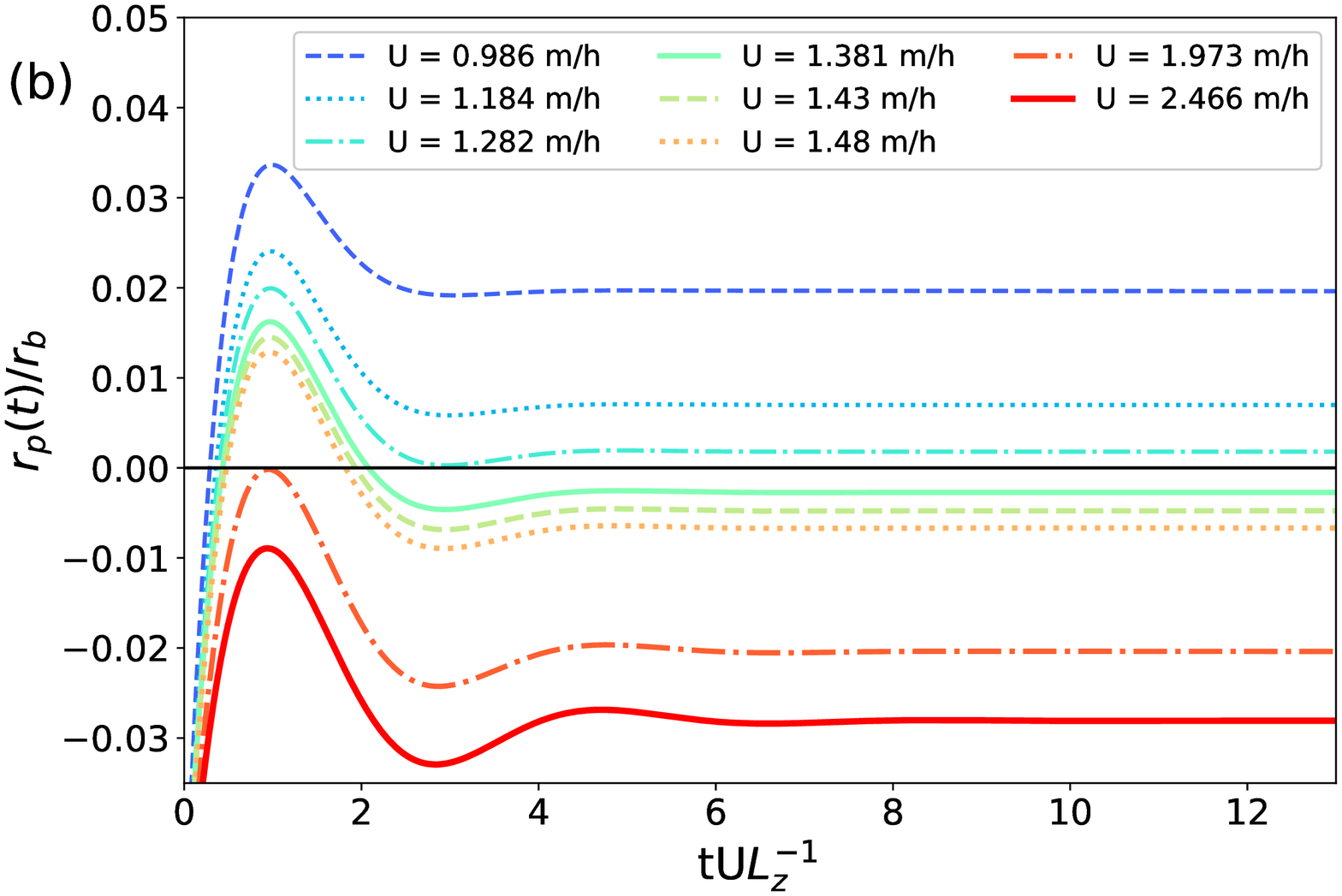}
\includegraphics[scale = 0.42]{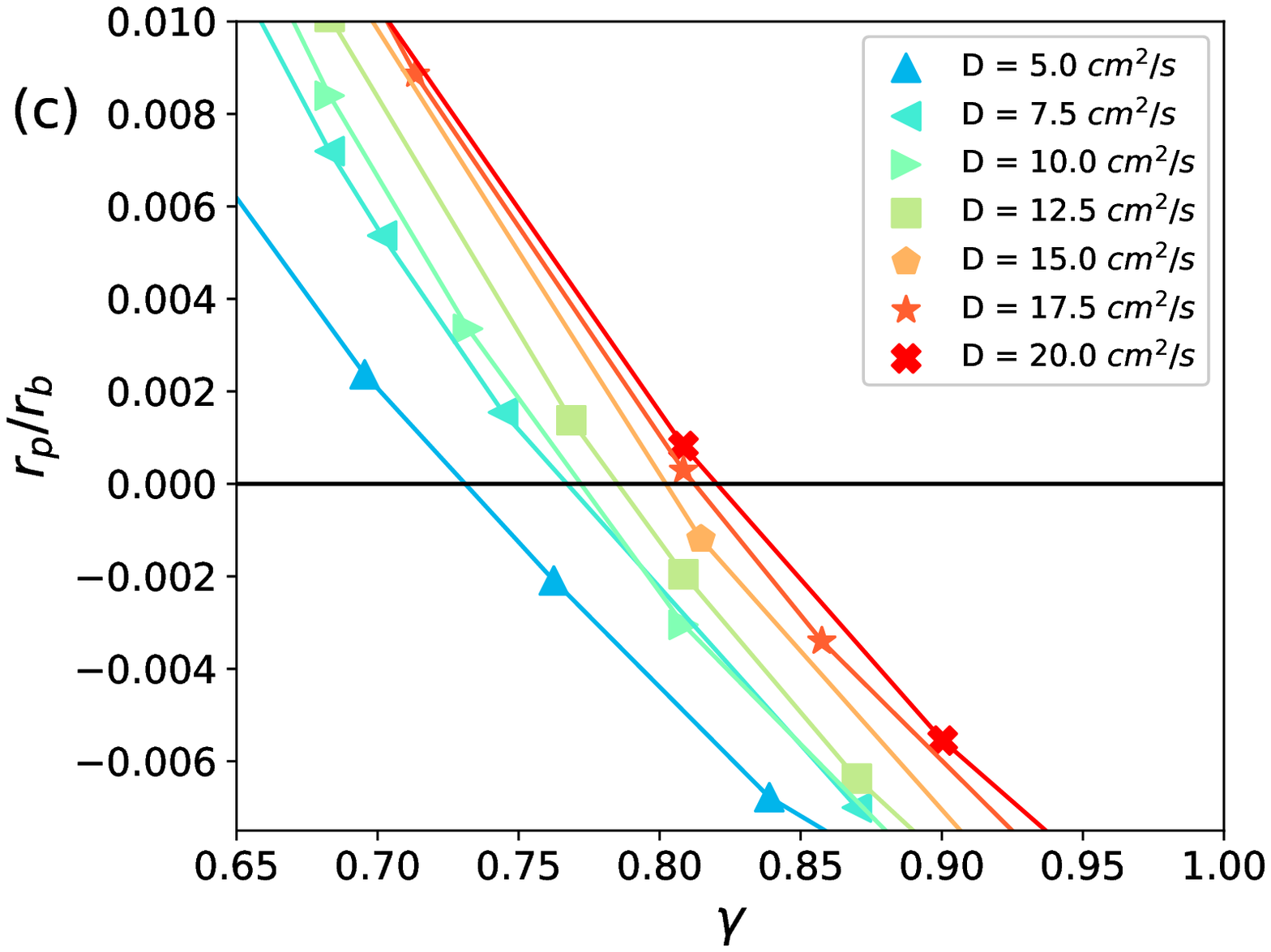}
\caption{\label{fig:patch_rp_gamma} 
(a) Average biomass density, on a logarithmic scale, versus time, for $D=20$~cm$^2$~s$^{-1}$, $L_z=60$~m and different values of the advection intensity $U$ in the steady-flow case. Vertical lines indicate $t = 2\pi L_{z}/U$, the time of one flow roll revolution. 
(b) Population per-capita growth rate $r_p(t)$, normalized by the intrinsic net growth rate $r_b$, as a function of time, normalized by the advective timescale $L_z/U$ of the steady-flow case, for various values of $U$, $D=20$~cm$^2$~s$^{-1}$ and $L_z=60$~m. 
(c) Per-capita growth rate $r_p$ (constant value attained after a transient), normalized by the intrinsic net growth rate $r_b$, versus the ratio of reactive to advective timescales $\gamma$, in the steady-flow case, for different values of $D$. 
}
\end{figure}

The flow intensity $U$ is then varied in a broad range to examine possible changes of behavior due to advection by the coherent flow. We particularly focus on the upper bound of the turbulence window (taking $5$~cm$^2$~s$^{-1} \leq$ $D \leq 20$~cm$^2$~s$^{-1}$), for which numerical simulations 
reveal more useful. 
The effect of increasing $U$ for fixed $D$ is apparent in Fig.~\ref{fig:patch_rp_gamma}(a), showing $\langle \theta \rangle$ as a function of time. These results were obtained using a localized initial condition corresponding to a small patch of population density located in the central upper part of the domain, close to $(x=L_{x}/2, z=0$), but we verified that the overall phenomenology stays unchanged when considering a uniformly spread initial population. The coherent flow reduces the growth of $\langle \theta \rangle$ and eventually causes an extinction when its intensity is large enough. 
The growing or decaying temporal behavior is already quite well established after one large eddy turnover time [see vertical lines in Fig.~\ref{fig:patch_rp_gamma}(a)], 
here estimated as $2\pi L_{z}/U$, approximating streamlines with perfectly circular orbits of radius $L_z$.  At later times, the average biomass density 
continues to grow exponentially at a constant rate. 
%%%%%%%%%%%%%%%%%%
% FIG. 3
%%%%%%%%%%%%%%%%%%

In order to characterize the bloom to no-bloom transition induced by advection, we measure the per-capita growth rate $r_p(t)$ [see Eq.~(\ref{eq:pc_gr})]. This quantity, normalized by the intrinsic total (birth minus death) growth rate at the surface $r_b=I_{in}/(H+I_{in})p_{max}-l$, versus time normalized by $L_z/U$, is shown in Fig.~\ref{fig:patch_rp_gamma}(b). Here, a uniform initial population density was chosen. As it can be seen, at large enough times, for all $U$, $r_p(t)$ approaches a constant value $r_p$, confirming the exponential character of growth or decay of $\langle \theta \rangle$. Furthermore, the large-time 
value $r_p$ decreases from positive values (for low $U$) to negative ones (at larger $U$), therefore allowing a robust estimate of the critical flow intensity at the transition.

As first indicated in~\cite{Abraham1998}, where horizontal patchiness was numerically studied adopting an NPZ (for nutrient-phytoplankton-zooplankton) model in a turbulent flow, we expect that also in the present case the dynamics are primarily controlled by the interplay between advection and reaction mechanisms. To quantify the relative weight of the latter processes, we consider the ratio of the biological timescale $r_b^{-1}$ to the flow timescale $L_z/U$, i.e.:
\begin{equation}
    \gamma = \frac{U}{r_b L_{z}}.
    \label{eq:gamma}
\end{equation}
Figure~\ref{fig:patch_rp_gamma}(c) reports the 
(constant) per-capita growth rate $r_p$ as a function of $\gamma$. From this plot, one can clearly see that the survival/extinction transition caused by the flow occurs for $\gamma=O(1)$, in correspondence with $r_p$ turning from positive to negative. Essentially, a bloom can take place ($r_p>0$) when the biological growth is faster than the advective transport ($\gamma<1$) to the less favorable deeper part of the domain. The proximity of the data obtained with different values of $D$ highlights the generality of this mechanism and confirms the weak effect of the turbulent diffusivity in this picture. We remark that we could not detect a transition to a no-bloom regime for $D<5$~cm$^2$~s$^{-1}$, even with very large values of $U$.

Further insight comes from inspection of the spatial structure of the population density field $\theta(x,z,t)/\langle \theta \rangle$ (normalized with $\langle \theta \rangle$) at a given time (Fig.~\ref{snapfieldsteady}). While in the absence of flow the population is uniformly distributed along the horizontal and decreases with depth, nonzero advection causes an increase of $\theta$ in the downwelling region (at $x=L_{x}/2$).  This feature gets accentuated by increasing $U$, with the population accumulating in thinner and thinner filaments outside vortices, and particularly in the one located at $x=L_{x}/2$. 
Such a behavior points to the relevance of strain-dominated flow regions for the spatial organization of the population and the formation of fine structures. In our flow, as it can also be easily seen in Figs.~\ref{snapfieldsteady}(b,c), the latter regions are close to the hyperbolic points corresponding to the vertices of the squares of side $L_z$ containing the rolls. 
Among such points, clearly, a prominent role is played by the point $(x,z)=(L_x/2,0)$, where the flow locally compresses the scalar field $\theta$ along the $x$-axis (and stretches it in the $z$-direction), in the region of highest growth rate (i.e. at the surface).

Relying on the above picture, a useful interpretation of the dynamics observed in our simulations is offered by an appropriate adaptation of the plankton filament model~\cite{Martin2000}, originally introduced to describe the formation of fine structures in 2D flows. To apply this reasoning, we neglect the sinking speed, which is considerably smaller than the advecting velocity close to the transition to extinction, as well as self-shading, as close to an extinction the population density is low everywhere and because our main point of interest is at the surface. Under these hypotheses, Eq.~(\ref{full}) becomes
\begin{equation}
\partial_t \theta + \bm{u} \cdot \bm{\nabla} \theta = \left( \frac{I_{in}}{H+I_{in}} \, p_{max} \, e^{-\kappa_{bg}z} -l \right) \theta +
D \bm{\nabla}^2 \theta.
\label{eq:filament_model_2d}
\end{equation}
Since, as argued above, we are interested in the dynamics at the surface, close to the hyperbolic point at $x=L_{x}/2$, the net growth rate will be $r_b$. 
Moreover, we can 
write the population density as: 
\begin{equation}
\theta(x,z,t)=\theta_{back} \, e^{r_b t} + \theta'(x,z,t), 
\label{eq:filament_model_1d_dec}
\end{equation}
where $\theta_{back}$ is a background population density and $\theta'$ represents the  perturbation determined by the flow.
Using this decomposition, it is not difficult to see that the equation governing the dynamics of $\theta'$ is the same as Eq.~(\ref{eq:filament_model_2d}).
Following~\cite{Martin2000}, we can then consider only the 1D dynamics for the population fluctuation in the compressing (or cross-filament) direction, because along the filament $\theta'$ should vary less due to the stretching operated by the flow. In this region, the flow can be locally approximated as $\bm{u}=\left(-\lambda (x-L_{x}/2),\lambda z\right)$, where $\lambda=kU \approx U/L_z$ is the strain rate. 
Therefore, from Eq.~(\ref{eq:filament_model_2d}), for the cross-filament dynamics one has: 
\begin{equation}
\partial_t \theta' - \lambda \left(x-L_{x}/2\right) \partial_x \theta' = r_b \theta' + D \partial_x^2 \theta'.
\label{eq:filament_model_1d}
\end{equation}
The solution of the above equation is (see also~\cite{Martin2000}):
\begin{equation}
\theta'(x,t) = \theta'_0 \, e^{-\frac{\left(x-L_{x}/2\right)^2 \lambda}{2D}} \, e^{(r_b-\lambda)t},
\label{eq:filament_solution}
\end{equation}
with $\theta'_0$ a constant. From this expression we can see that, in the $x$~direction, the population density field keeps the same (Gaussian) shape at different times. The filament width $\sigma=\sqrt{D/\lambda}$ does not depend on time and is only determined by the physical parameters associated with fluid transport. As the flow intensity increases, so does the strain rate, which explains the thinning of filaments and the more important localization of the population for higher values of $U$. Growth or decay over time, instead, depends on whether $r_b$ is larger or smaller than $\lambda$, respectively. This simple model thus provides theoretical support to the survival/extinction criterion based on the ratio between the biological and flow timescales, $\gamma$.

The above model accounts for the dynamics at the surface and, strictly speaking, 
it is only there that its predictions should apply. If the population cannot survive at the surface, however, it should not deeper below either, due to the reduced growth rate, which makes the conclusion appear more general. Considering that, differently from the 1D filament case, in our fully 2D model both the strain rate and the growth rate vary with depth, and that sinking and self-shading might also play a minor role, the comparison between our previous estimate of the control parameter, $\gamma=U/(L_z \, r_b)$, and that from Eq.~(\ref{eq:filament_solution}), $\lambda/r_b$, seems to us reasonable also from a quantitative point of view. Regarding the dependence on the vertical coordinate, we further note that the biological growth rate monotonously decays with $z$, and that the strain rate, in absolute value, decreases until half the total depth, before growing again in the lower half of the domain, but now acting in the opposite way (stretching instead of compressing the scalar in the $x$-direction). The combination of these effects, impacting both the width and the intensity of the filament, can then explain, in a qualitative way, 
the tendency, particularly visible in  Fig.~\ref{snapfieldsteady}(b), of this localized downwelling structure to fade around $z=L_z/2$.

\begin{figure}[h!]
\includegraphics[scale = 0.5]{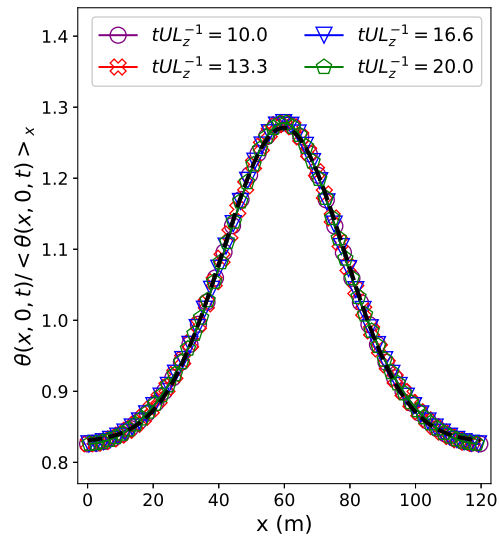}
\caption{
\label{fig:hprof}
Horizontal profiles of surface population density at different times (in units of the advective timescale $L_z/U$), normalized by their average values, $\theta(x,0,t)/\langle \theta(x,0,t) \rangle_x$, for the steady-flow case ($\Psi = \Psi_L^{st}$) with $U = 1.001$~m~h$^{-1}$ and $D = 20$~cm$^{2}$~s$^{-1}$. 
The dashed black line represents the Gaussian solution of Eq.~(\ref{eq:filament_model_2d}), $\Theta + \theta_i' \exp{\left[(x-L_{x}/2)^2/(2\sigma_{numerical}^2)\right]}$, with $\Theta=\theta_{back} \exp{(r_b t)}$ and $\theta_i' = \theta_0' \exp{[(r_b-\lambda)t]}$ [see also Eqs.~(\ref{eq:filament_model_1d_dec}-\ref{eq:filament_model_1d})]. The constants $\Theta$, $\theta_i'$ and $\sigma_{numerical}$ are fitting parameters.
}
\end{figure}

\begin{figure}[h!]
\includegraphics[scale = 0.57]{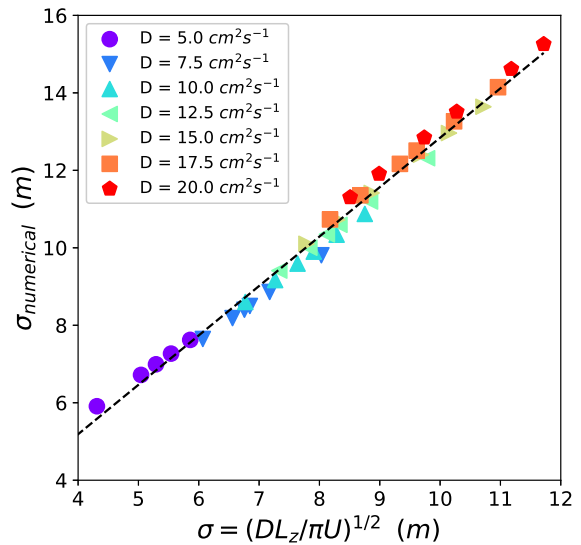}
\caption{
\label{gausfit} 
Filament width, estimated from a fit (in the interval $50$~m~$<x<70$~m) with a Gaussian function of horizontal profiles of population density from simulations with different values of $D$ and $U$, versus its theoretical prediction  $\sigma=\sqrt{D/(kU)}$, with $k=\pi/L_z$.  The dashed black line corresponds to $\sigma_{numerical}=0.085 + 1.276 \sigma$. 
}
\end{figure}

To test the validity of the above argument for our system, we examined the horizontal profiles of population density at $z=0$ from simulations with different values of $D$ and $U$, once $r_p(t)$ 
reached the constant value $r_p$. We found that such profiles are to good extent time independent and that their shape is well described by a Gaussian function.
Figure~\ref{fig:hprof} shows an example of the latter profiles $\theta(x,0,t)$, at different instants of time (for given values of $U$ and $D$), 
normalized by the corresponding average values $\langle \theta(x,0,t) \rangle_x$. By means of a fit in a subregion centered around $x=L_{x}/2$, where the phytoplankton patch is mainly localized, we then estimated the standard deviation of the Gaussian curves, $\sigma_{numerical}$, which provides a measure of the filament width $\sigma$. The results are compared to the theoretical prediction in Fig.~\ref{gausfit}, which indicates a strong correlation between the numerical and theoretical estimations of $\sigma$. As one can observe in the figure, we actually detect a tendency of the numerically estimated $\sigma$ to grow slightly faster than the theoretical one. However, such a small difference seems quite reasonable, taking into account the assumptions made for the theoretical prediction with respect to the details of our numerical setup. 
Note, too, that while the linear proportionality between $\sigma_{numerical}$ and $\sigma$ is quite robust, particularly for large values of $D$, the quality of the agreement (between the numerical and theoretical values) depends on the width of the central region chosen for the estimation of $\sigma_{numerical}$. 
%%%%%%%%%%%%%%%%%%
% FIG. 4
%%%%%%%%%%%%%%%%%%
% FIG. 5
%%%%%%%%%%%%%%%%%%

It is worth remarking that at the bottom, due to the similar structure of the deep and surface flow, the spatial organization of the reactive scalar parallels that found at the surface. An analogue reasoning in the straining regions close to $x=0$ or $x=L_{x}$ would always give extinction locally, however, as the prefactor in the exponent of the exponential involving time would be $-l-\lambda<0$, as the growth rate is negligibly small there. Hence, the relatively high values of population density at the bottom appear to be due to fluid transport (including sinking) only and the zero-flux boundary conditions.

Finally, according to Eq.~(\ref{eq:filament_solution}), in the limit of very small diffusivity, the filament width approaches zero while its density amplitude grows exponentially. Consequently, it becomes more and more difficult to observe an advection-driven extinction. From a practical point of view, this is essentially impossible in numerical simulations, as it would require an infinite spatial resolution, in order to resolve the cross-filament structure. These are likely the reasons why we could not detect the transition to no-bloom at sufficiently small values of $D$.

%%%%%%%%%%%%%%%%%%%%%%%%%%%%%%%%%%%%%%%%%%%%%%%%%%
\subsection{\label{sec:results cell} Unsteady flow}

\begin{figure}[h]
\includegraphics[scale = 0.58]{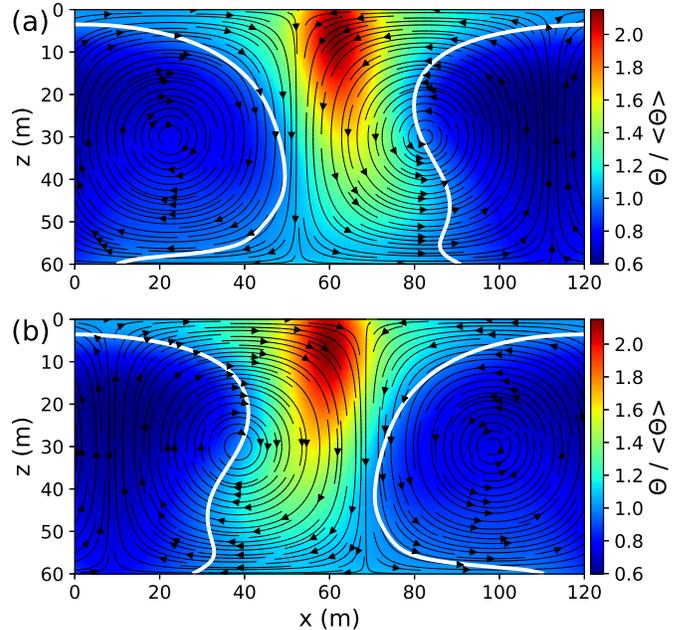}
\caption{
\label{snapfieldtimdep} 
Instantaneous population density field, normalized by its spatial average, $\theta(x,z,t)/\langle \theta \rangle$, in the unsteady-flow case, for $U = 1.4$~m~h$^{-1}$ and  $D=5$~cm$^2$~s$^{-1}$. The two panels correspond to two different times. The white line is the isoline $\theta/\langle \theta \rangle=1$ and the black lines are the streamlines of the flow field at the considered instants of time, with arrows indicating the circulation direction.
}
\end{figure}

We now consider a time-dependent large-scale flow, by allowing for lateral oscillations of the flow pattern
adopted in the previous section, which is enough to produce chaotic Lagrangian trajectories of fluid particles~\cite{vulpiani2009chaos,neufeld2009chemical}. More explicitly, the flow field will now be specified by Eq.~(\ref{flowlarge}), i.e. Eq.~(\ref{flowcomplt}) with $\Psi_{s} = 0$. The amplitude and pulsation of the roll oscillation are respectively set to $s=L_z/5$ and $\omega=\pi U/L_z$, corresponding to a fraction of the roll size and a period comparable to the advective timescale $L_z/U$, a choice that has been shown to be optimal to enhance chaotic diffusion~\cite{solomon1988chaotic,vulpiani2009chaos,lacorata2017chaotic,lacorata2019fsle}. Note that we do not allow for vertical oscillations, in order to keep the top and bottom boundaries of our domain at fixed vertical positions. Figure~\ref{snapfieldtimdep} shows two snapshots of the population field at different times in the constant per-capita growth-rate regime. These visualizations suggest that the dynamics are fairly similar to those in the stationary-flow case, although horizontal symmetry is now broken due to the lateral oscillations of the flow.
%%%%%%%%%%%%%%%%%%
% FIG. 6
%%%%%%%%%%%%%%%%%%

\begin{figure}[htb]
\includegraphics[scale = 0.5]{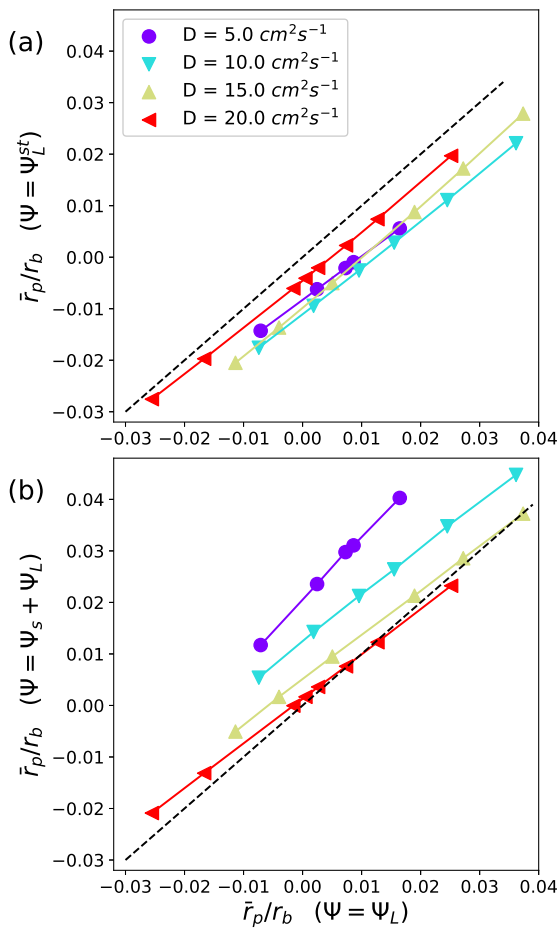}
\caption{\label{fig:7rpx1rp_St} 
(a) Time averaged normalized per-capita growth rate $\bar{r}_{p}/r_{b}$ of the steady-flow case vs the corresponding quantity from the unsteady-flow case. (b) Same as in (a) but for $\bar{r}_{p}/r_{b}$ from the multiscale flow case vs $\bar{r}_{p}/r_{b}$ from the unsteady-flow case.
In both (a) and (b), several values of 
the large-scale flow intensity ($0.79$~m~h$^{-1} \leq U_1 \leq 3.69$~m~h$^{-1}$) and of $D$ are considered. Fitting the data corresponding to a given value of $D$ [$(5,10,15,20)$~cm$^2$~s$^{-1}$] with a linear function, we obtain slopes that are always quite close to 1, particularly in (a) (slopes between 0.84 and 0.93); the data in (b) display a little more variability (with fitted slopes between 0.86 and 1.21). The black dashed lines have unitary slope.
}
\end{figure}

To confirm this observation we performed the same analysis as in Sec.~\ref{sec:results steady}. The results indicate that the overall phenomenology remains unchanged, with only little quantitative differences. The temporal behaviors of both the average biomass density $\langle \theta \rangle(t)$ and the per-capita growth rate $r_p(t)$ are similar to those observed with the steady flow 
[Figs.~\ref{fig:patch_rp_gamma}(a,b), respectively], 
but they now present small oscillations with a frequency corresponding to that of the roll lateral displacement (not shown). 
As for the critical advection intensity $U_c$ determining the bloom/no-bloom transition, it is found to be slightly higher in the present time-dependent case. The increase with respect to the previous, steady, case depends on the value of the small-scale diffusivity (about $6\%$ for $D = 20$~cm$^2$~s$^{-1}$ and $18\%$ for $D = 10$~cm$^2$~s$^{-1}$), but the dependency of $U_c$ on $D$ remains weak. Considering that the explicit time dependency of $\Psi$ in Eq.~(\ref{flowlarge}) now gives rise to chaotic diffusion of Lagrangian particles, and hence to an effective diffusivity larger than $D$, such an increase of $U_c$ seems to us reasonable, from a qualitative point of view.
A more quantitative assessment of the comparison between the unsteady and steady flow cases is illustrated in Fig.~\ref{fig:7rpx1rp_St}(a). Here we  show $\bar{r}_{p}/r_{b}$ of the steady flow case as a function of $\bar{r}_{p}/r_{b}$ in the unsteady case, for several values of $U$ and $D$. As it can be seen, over the range of values of $D$ and $U$ explored, the two quantities are almost perfectly correlated, corroborating the idea that the lateral oscillations do not produce any major modifications.
%%%%%%%%%%%%%%%%%%
% FIG. 7
%%%%%%%%%%%%%%%%%%

%%%%%%%%%%%%%%%%%%%%%%%%%%%%%%%%%%%%%%%%%%%%%%%%%%%%%
\subsection{\label{sec:results multiscale} Multiscale flow}

We now extend our analysis to a multiscale flow, mimicking a turbulent one, specified by the full streamfunction in Eq.~(\ref{flowcomplt}). 
Again, the flow is explicitly time dependent and performs oscillations, now on different scales, with amplitudes $s_{i} = L_{i}/10$ and pulsations $\omega_i = \pi U_{i}/L_{i}$ (a choice that is analogous to that of Sec.~\ref{sec:results cell}, considering that $L_{1} = L_{x}$). The first, and largest-scale, mode only oscillates laterally (as in the previous section) while smaller-scale flow components are allowed to move also in the vertical direction. Close to the vertical boundaries, however, their oscillations are damped according to Eq.~(\ref{eq:dump}), in order to respect no-flux boundary conditions for the reactive scalar.
We choose a number of modes that allows spanning the scale range going from the domain size $L_1=120$~m to the smallest length scale $L_{n_{k}}=1$~m, corresponding to $\approx 1/(5 \kappa_{bg})$, where $\kappa_{bg}^{-1}$ is related to the growth dynamics, as it is the typical length over which light is absorbed by the medium. Such small length scale also roughly corresponds to the scale that can be estimated from Richardson scaling of diffusivity with length, $\ell \sim (2/3)^{3/4} \epsilon^{-1/4} D(\ell)^{3/4}$ \cite{lacorata2017chaotic,boffetta2002relative}, using the values of diffusivity explored in the previous sections, $5$~cm$^2$~s$^{-1}<D<20$~cm$^2$~s$^{-1}$, and values of the kinetic energy dissipation rate $\epsilon \approx (10^{-8}-10^{-6})$~m$^2$~s$^{-3}$ that appear reasonable for oceanic turbulence~\cite{kiorboe1995planktivorous,barton2014impact,lindemann2017dynamics}. We then set the scale separation factor to $\eta = 2$ and the number of modes to $n_{k} = 7$. Finally, we assume a Kolmogorov scaling of velocity, $U_i=U_1(L_i/L_1)^{1/3}$. 

\begin{figure}[h]
\includegraphics[scale = 0.58]{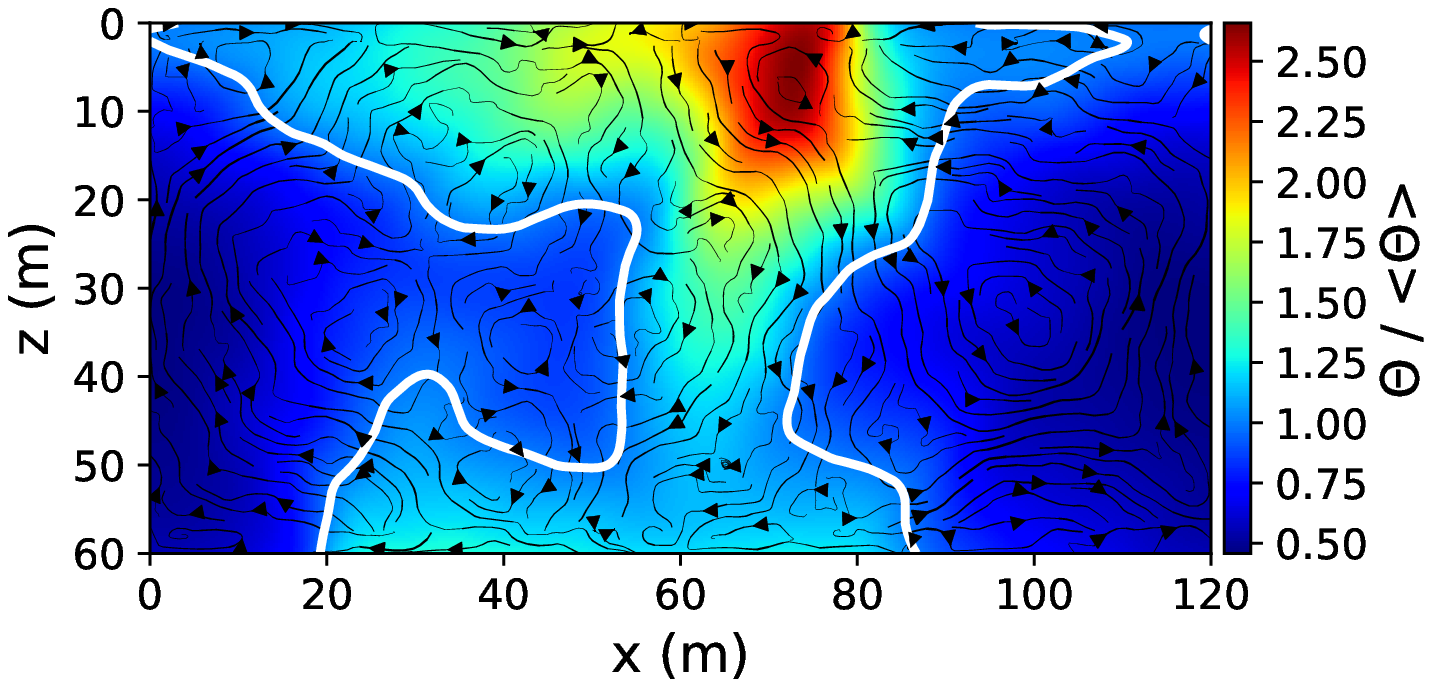}
\caption{
\label{fig:snapfield7mod} 
Instantaneous population density field, normalized by its spatial average, $\theta(x,z,t)/\langle \theta \rangle$, in the multiscale-flow case, for $U = 1.021$~m~h$^{-1}$ and $D=5$~cm$^2$~s$^{-1}$, in the  regime $\overline{r}_p(t)=\mathrm{const}$. The white line is the isoline $\theta/\langle \theta \rangle=1$ and the black lines are the streamlines of the flow field at the considered instant of time, with arrows indicating the circulation direction.
}
\end{figure}

Figure~\ref{fig:snapfield7mod} presents the population density field at a given time ($t^*=520$~h), normalized by its spatial average. As in previous visualizations, we select the time $t^*$ such that the dynamics have already reached the 
constant growth-rate regime characterized by $\overline{r}_p(t)=\mathrm{const}$. The iso-contours of the streamfunction at the same time (black lines in the figure), allow to appreciate the presence of eddies of different sizes and the more disordered spatial structure of the velocity field. Although the latter small-scale features reflect in the spatial distribution of the population, which is now irregular, the signature of the largest-scale flow is still apparent, particularly in the $\theta$ patch at the center of the domain ($x \simeq L_x/2=60$~m) and close to the surface. 
%%%%%%%%%%%%%%%%%%
% FIG. 8
%%%%%%%%%%%%%%%%%%

It is interesting to compare the growth rate $\overline{r}_{p}/r_{b}$ measured in this multiscale setting and in the previous ones, e.g. in the one-mode unsteady-flow case. As it can be seen in Fig.~\ref{fig:7rpx1rp_St}(b), the estimates from the two cases are still quite correlated, and diffusivity plays a rather weak role [similarly to the previous comparison, reported in Fig.~\ref{fig:7rpx1rp_St}(a)]. Still, we notice that in Fig.~\ref{fig:7rpx1rp_St}(b), for small enough $D$, it is possible to observe that the multiscale estimate of the growth rate $\overline{r}_p$ tends to be slightly larger than its counterpart in the absence of small eddies. We then argue that the latter flow features provide an effect that partially compensates the deadly action of the large-scale coherent flow, but that is only measurable for $D \leq 10$~cm$^2$~s$^{-1}$.
 
\begin{figure}[h]
\includegraphics[scale = 0.37]{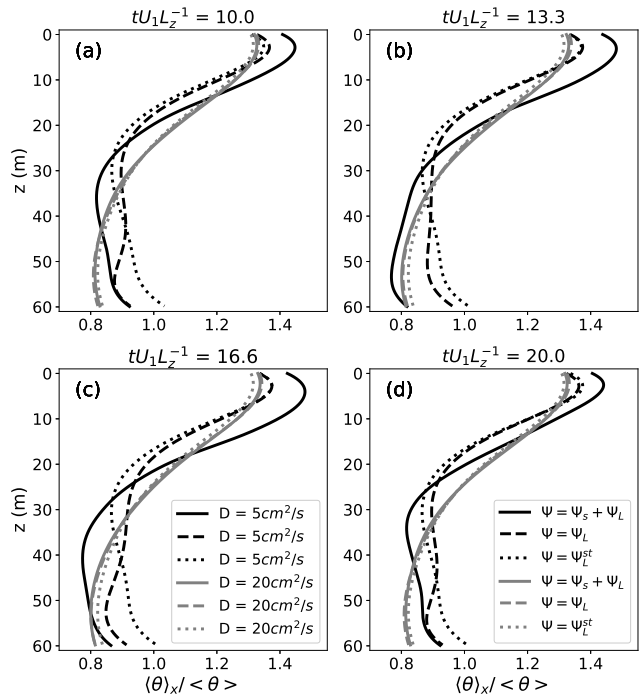}
\caption{
\label{fig:denscomp} 
Vertical population density profiles $\langle \theta \rangle_x$, normalized by the global spatial average $\langle \theta \rangle$, for the different streamfunctions $\Psi$ considered, $U_{1} = 1.001$~m~h$^{-1}$ and  $D=(5,20)$~cm$^2$~s$^{-1}$. Different line types correspond to different choices of $\Psi$, with black and gray curves indicating the different values of $D$. 
Panels (a-d) correspond to different instants of time, in the constant growth-rate regime $\overline{r}_p(t)=\mathrm{const}$, as specified in the plot titles (in units of the advective timescale $L_z/U_1$).
}
\end{figure}

To further investigate the impact of small-scale fluid motions, we also analyze vertical profiles $\langle \theta \rangle_x(z)$ of the phytoplankton distribution (similarly to what is done in~\cite{taylor2011shutdown,huisman2002population,ryabov2010vertical}), obtained by averaging $\theta(x,z,t)$ over the horizontal coordinate $x$ at fixed instants of time. Such profiles, normalized by the corresponding global spatial averages $\langle \theta \rangle$, are shown in Fig.~\ref{fig:denscomp} for all the flow cases studied (one-mode steady-flow, one-mode oscillatory flow, multiscale time-dependent flow) at common given times. Independently of the considered flow or value of $D$, their shape is always characterized by a maximum at small, but finite, depth and a decrease deeper below the surface, plus a second inflection point close to the bottom boundary. These features are typical for sinking phytoplankton species~\cite{huisman2002sinking}, whereas non-sinking ones would display a maximum at the surface~\cite{huisman1999critical}.
%%%%%%%%%%%%%%%%%%
% FIG. 9
%%%%%%%%%%%%%%%%%%

The similarity of the profiles obtained in different configurations (Fig.~\ref{fig:denscomp}) points to the dominance of advection by the large-scale coherent flow, as including its time dependence or smaller scales does not alter the general picture substantially. Note, however, that for sufficiently low $D$, the addition of small scales favors, to small but measurable extent, a localization of the population close to the surface, akin to the vertically nonhomogeneous distribution typical of the no-flow case [see Fig.~\ref{snapfieldsteady}(a)] and increased possibility of survival. 

\begin{figure}[h]
\includegraphics[scale = 0.37]{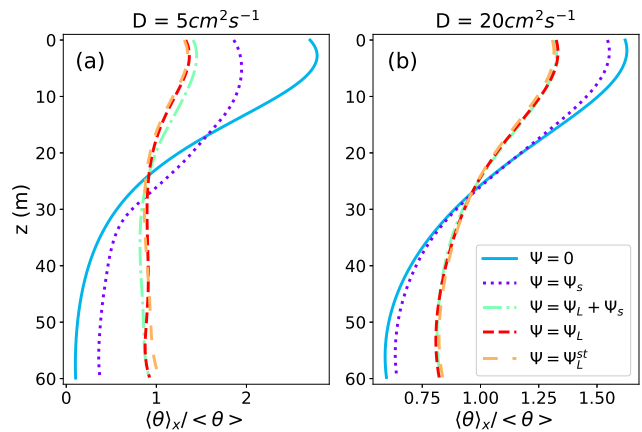}
\caption{
\label{fig:densallcases} 
Normalized vertical population density profiles $\langle \theta \rangle_x/\langle \theta \rangle$ for $\Psi = 0$ (no flow), $\Psi=\Psi_L^{st}$ (large-scale steady flow), $\Psi=\Psi_L$ (large-scale time-dependent flow), $\Psi=\Psi_L+\Psi_s$ (multiscale time-dependent flow), $\Psi=\Psi_s$ (small-scale time-dependent flow, without $\Psi_L$) and $U_{1} = 1.001$~m~h$^{-1}$. Panels (a) and (b)  respectively refer to $D=5$~cm$^2$~s$^{-1}$ and $D=20$~cm$^2$~s$^{-1}$. Note the different value ranges on the horizontal axes in (a) and (b). All the profiles here reported are computed at a common fixed time $tU_1L_z^{-1}=10$, for which $\overline{r}_p(t)=\mathrm{const}$. 
}
\end{figure}

The importance of the large-scale flow can be even better appreciated by inspecting Fig.~\ref{fig:densallcases}. Here, again for a common fixed time ($t U_1 L_z^{-1}=10$) in the (statistically) constant growth-rate regime, we show the normalized vertical profiles $\langle \theta \rangle_x/\langle \theta \rangle$, for the smallest and largest value of diffusivity used [$D=5$~cm$^2$~s$^{-1}$ and $20$~cm$^2$~s$^{-1}$ in panels (a) and (b), respectively], for different flow types. Specifically, we examine the following different combinations: $\Psi=0$ (no flow), $\Psi=\Psi_L^{st}$ (large-scale steady flow), $\Psi=\Psi_L$ (large-scale time-dependent flow), $\Psi=\Psi_L+\Psi_s$ (multiscale time-dependent flow), $\Psi=\Psi_s$ (time-dependent flow without the large-scale contribution provided by $\Psi_L$). The last case was explicitly added to test the relevance of the large-scale advection. It is apparent that whenever $\Psi_L$ is present the population gets homogenized in the vertical direction, with respect to the no-flow case. The addition of time dependency and small scales to the flow turns out to play only a minor role, as the corresponding profiles are essentially indistinguishable from the one obtained with $\Psi_L$ only. When the latter contribution is removed and the flow only possesses smaller scales, instead, the population distribution retrieves the  vertically nonhomogeneous character typical of the $\Psi=0$ case. In such a case, in fact, the vertical profile $\langle \theta \rangle_x$ approaches the one obtained without flow, as it is particularly evident in Fig.~\ref{fig:densallcases}(b) (where $D=20$~cm$^2$~s$^{-1}$). 
Finally, it seems to us that Fig.~\ref{fig:densallcases} summarizes in an effective way the main outcome of this work, meaning the outstanding relevance of advection by the large-scale coherent flow, as the dominant mechanism controlling phytoplankton dynamics in the present setting. 
%%%%%%%%%%%%%%%%%%
% FIG. 1
%%%%%%%%%%%%%%%%%%

%%%%%%%%%%%%%%%%%%%%%%%%%%%%%%%%%%%%%%%%%%%%%%%%%%%%%
\section{\label{sec:conclus} Conclusions}

We numerically investigated the dynamics of sinking phytoplankton in a stirred 2D fluid layer where the vertically decreasing light availability is the only limiting factor for biological growth. For this purpose we extended a previous theoretical 1D model~\cite{huisman1999critical,ebert2001critical,huisman2002sinking}, where turbulent motions were only described in terms of an effective diffusivity, by taking into account in an explicit way the transport operated by a structured fluid flow. The choice to neglect possible heterogeneities in the nutrient distribution was motivated by our goal to focus on the role of transport mechanisms. While clearly this poses some limitations in relation to real natural environments, where nutrients can also affect biological growth, such a configuration still appears reasonable for, nutrient-rich, eutrophic habitats, namely shallow warm lakes or high-latitude oceans.

A major outcome of the simplified theoretical model~\cite{huisman2002sinking} mentioned above 
was to provide evidence of two transitions between extinction and survival of the population, depending on the turbulent intensity (for deep enough fluid layers). Our aim, here, was to explore the impact of a more realistic representation of the advecting velocity field on the survival-to-extinction transition, for which no analytical prediction is available, occurring at large turbulent intensity when biological production cannot compensate turbulent mixing to sustain the bloom in the well-lit region close to the surface. Using realistic parameter values for the biological dynamics~\cite{huisman2002sinking}, we then considered a domain with a fixed depth representative of the mixed layer, in the presence of flows of progressively increasing complexity, relying on a kinematic-flow approach~\cite{lacorata2017chaotic}. We first examined a velocity field possessing a single large-scale stationary mode, in the form of two recirculating cells spanning the horizontal extent of the system~\cite{stommel1949trajectories,solomon1988chaotic}. Such a spatial structure was intended to mimic the large-scale features observed in realistic flows, as those arising from buoyancy driven convection~\cite{solomon1988chaotic,vallis2017atmospheric} or wind-driven Langmuir circulation~\cite{stommel1949trajectories,okubo2001diffusion,denman1983time}. We then added time dependency in the form of lateral oscillations of such a flow pattern, and finally included spatially and temporally varying smaller scales.

Our results indicate that advection plays a relevant role on the biological dynamics. Indeed, persistent large-scale motions reduce the per-capita growth rate and can eventually lead to the suppression of the bloom, when the flow is intense enough. This effect is found  to be controlled by the ratio between the characteristic biological and flow timescales, similarly to what occurs for plankton horizontal dynamics stirred by mesoscale ocean eddies~\cite{mckiver2009plankton}. From a general perspective, a similar harmful role of the advecting flow was also put in evidence in previous LES of turbulent thermal convection~\cite{taylor2011shutdown}, and in a study considering a steady cellular flow and a matrix-based approach to compute the biological growth rate~\cite{lindemann2017dynamics}. However, those studies neglected the phytoplankton self-shading~\cite{lindemann2017dynamics} and also sinking~\cite{taylor2011shutdown}. Moreover, in both of them it is less straightforward than in our work to disentangle contributions from large and small flow scales, either because the latter are essentially absent~\cite{lindemann2017dynamics}, or because they dynamically interact with the large-scale ones~\cite{taylor2011shutdown}.  

The main finding of the present study is that the large-scale flow dominates the dynamics, which are only weakly affected by (temporally and/or spatially)  smaller-scale fluid motions. This is revealed by both the strong correlation found for the critical flow intensities (for the transition), and the similar vertical population profiles, in the different flow cases. Even in the  presence of a multiscale flow, the velocity field at the largest scale has a strong signature on the dynamics, as it drives the localization of the population in a patch at the center (with respect to the horizontal coordinate) of the domain. This is a region of phytoplankton downwelling (similarly to what is observed in Ref.~\cite{lindemann2017dynamics}), corresponding to the location of the straining point associated with the largest-scale flow mode (i.e.~the separatrix between the two largest rolls).

In the (large-scale) one-mode, steady, flow case, we have been able to rationalize the picture by adapting the plankton filament model originally introduced in Ref.~\cite{Martin2000}. This allowed us to provide a quantitative justification for the control exerted by the biological-to-flow timescale ratio on the transition to extinction. As shown by our analysis of the multiscale-flow case, the presence of smaller-scale fluid motions tends to partially disrupt the regular spatial distribution of the population due to the flow at large scale, and the associated central downwelling filament. 
This was further confirmed by the comparison of phytoplankton density vertical profiles in two multiscale flows, one of which does not possess the largest-scale mode. Indeed, in the strain region between the largest eddies, the combined action of the flow and of small-scale diffusivity vertically homogenizes the population, thus hindering survival. When only smaller eddies are present, however, the planktonic population localizes closer to the surface, and spreads more over the horizontal, giving rise to a situation resembling that of the no-flow case, which is less prone to extinction.

We hope that the analysis reported here can contribute to the understanding of the basic mechanisms controlling the interplay between fluid transport and phytoplankton growth dynamics. The favorable comparison of some of our results with those obtained in the framework of more realistic fluid models~\cite{taylor2011shutdown} seems to us interesting in light of parameterizations of plankton cycles in numerical models. Several extensions can be envisaged, in a rather natural way. On one side, it would be interesting to consider a three-dimensional 
setup for our kinematic flow~\cite{lanotte2016effects,lacorata20083d}, to explore possible links between the vertical organization of phytoplankton and its horizontal patchiness. On the other, we believe that accounting for vertical variations of the turbulent intensity could provide a more realistic representation of real aquatic environments under stirring. 

%%%%%%%%%%%%%%%%%%%%%%%%%%%%%%%%%%%%%%%%%%%%%%%%%%

\appendix*

%%%%%%%%%%%%%%%%%%%%%%%%%%%%%%%%%%%%%%%%%%%%%%%%%%

\section{\label{sec:appendixa} Numerical method}

The dynamics specified by Eq.~(\ref{full}) are numerically integrated by means of a pseudo-Lagrangian algorithm~\cite{abel2001front,berti2005mixing,sandulescu2007plankton,GF2020}, based on the splitting of the advection, reaction and diffusion terms. Advection by the full velocity $\bm{v}$ (including both the fluid flow and phytoplankton sinking) is integrated backwards in time, for each grid point on which the population density field $\theta(x,z,t)$ is defined. This allows to determine the origin of the Lagrangian trajectory ending at the considered grid point after a time step $dt$. The value of $\theta$ at such Lagrangian origin, which is generally not on the numerical grid, is then determined by bilinear interpolation using the values of the field on the nearest grid points. Once known, the latter value of $\theta$ is used as the initial condition for the forward integration of the reaction dynamics over a time step. Finally, the integration of the diffusion term is carried out by means of a finite-difference implementation~\cite{sandulescu2007plankton,GF2020}, using a smaller time step $dt_D=dt/10$, meaning that $10$ diffusive steps are performed after each advection and reaction integration over $dt$. The choice of the value of $dt_D$ results from the two conditions required by the method. On one side, the physical diffusion coefficient $D$ has to be larger than the numerical one,  $D_{n} \propto dx^{2}/dt$, with $dx$ being the mesh size. On the other side, the stability condition for the Eulerian diffusive step is  $Ddt_{D}/dx^{2} < 1$. In our case, this leads to the choices $dt = 0.01$~h for the time step, and $dx = O(0.1)$~m for the grid size, allowing to resolve the typical length scales of reaction, advection and diffusion processes, for the values of $U$ and $D$ adopted.

%\bibliography{apssamp}

\providecommand{\noopsort}[1]{}\providecommand{\singleletter}[1]{#1}%

\end{document}